\documentclass[fleqn,10pt,onecolumn]{wlscirep}
\usepackage[utf8]{inputenc}
\usepackage[T1]{fontenc}

\newcommand{\kms}{\,{\rm km \, s^{-1}}}

\newcommand{\kpc}{\,{\rm kpc}}

\newcommand{\oversim}[2]{\protect{\mbox{\lower0.5ex\vbox{%
   \baselineskip=0pt\lineskip=0.2ex
   \ialign{$\mathsurround=0pt #1\hfil##\hfil$\crcr#2\crcr\sim\crcr}}}}} 
\newcommand{\bb}[1]{\ifmmode \mbox{\boldmath $ #1$} \else  \mbox{\boldmath $#1$} \fi}
\catcode`\"=\active\let"=\"  

\def\3{{\ss} }

\def\c12{{1\over 2}}

\def\d{{\rm d}}   
   
\def\plusplus{\raise 0.3ex\hbox{${\scriptstyle ++}$}{}}

\def\and{{{\rm M}31}}
\def\mw{{\rm MW}}
\def\lmc{{\rm LMC}}
\def\smc{{\rm SMC}}

\def\yr{\,{\rm yr}}
\def\d{{\rm d}\,}

\def\apj{Astrophys. J.}
\def\mnras{Mon. Not. R. Astron. Soc.}
\def\aap{Astron. \& Astrophys.}
\def\araa{Ann. R. Astron. \& Astrophys.}
\def\aj{Astron. J.} 
\def\jcap{J. Cosmol. Astropart. Phys.}

\def\apjs{Astrophys. J. Supp.}
\def\procspie{Soc. Photo. Instr. Engin.}
\def\apjl{Astrophys. J. Lett.}

\renewcommand{\figurename}{Fig.}

\title{Detection of the Milky Way reflex motion due to the Large Magellanic Cloud infall}

\author[1,*,+]{Michael S. Petersen}
\author[1,2,+]{Jorge Pe\~narrubia}

\affil[1]{Institute for Astronomy, University of Edinburgh, Royal Observatory, Blackford Hill, Edinburgh EH9 3HJ, UK}
\affil[2]{Centre for Statistics, University of Edinburgh, School of Mathematics, Edinburgh EH9 3FD, UK }

\affil[*]{michael.petersen@roe.ac.uk}

\affil[+]{these authors contributed equally to this work}

\begin{document}

\thispagestyle{empty}

\maketitle

\vspace{-7pt}

\section*{Main}
\textbf{
\noindent 
The Large Magellanic Cloud is the most massive satellite galaxy of the Milky Way, with an estimated mass exceeding a tenth of the mass of the Milky Way\cite{besla10,boylankolchin11,penarrubia16,shao18,erkal19}. Just past its closest approach of about 50 kpc, and flying by the Milky Way at an astonishing speed of $327\kms$ (ref.\cite{kallivayalil13}), the Large Magellanic Cloud can affect our Galaxy in a number of ways, including dislodging the Milky Way disc from the Galactic centre-of-mass\cite{gomez15,garavitocamargo19,petersen20a}. Here, we report evidence that the Milky Way disc is moving with respect to stellar tracers in the outer halo ($40<r<120\kpc$) at $v_{\rm travel}=32^{+4}_{-4}\kms$, in the direction $(\ell,b)_{\rm apex}=(56^{+9}_{-9},-34^{+10}_{-9})$ degrees, which points at an earlier location on the LMC trajectory. The resulting reflex motion is detected in the kinematics of outer halo stars and Milky Way satellite galaxies with accurate distances, proper motions and line-of-sight velocities. Our results indicate that dynamical models of our Galaxy cannot neglect gravitational perturbations induced by the Large Magellanic Cloud infall, nor can observations of the stellar halo be treated in a reference frame that does not correct for disc reflex motion.
Future spectroscopic surveys of the stellar halo combined with Gaia astrometry will allow for sophisticated modelling of the Large Magellanic Cloud trajectory across the Milky Way, constraining the dark matter distribution in both galaxies with unprecedented detail.
}

Numerical simulations that follow the accretion of a massive Large Magellanic Cloud (LMC) into our Galaxy predict that the MW disc will be displaced from the Galactic centre of mass with a speed of $v_{\rm travel}\approx v_\lmc (M_\lmc/M_\mw)$ (refs.\cite{gomez15, garavitocamargo19, petersen20a}). If true, the disc motion should be detectable in the apparent kinematics of halo stars with long dynamical times as a simple dipole \cite{petersen20a}. The peak of the blue-shifted velocities, or {\it apex}, is the direction to which the MW disc is currently travelling. Here, we measure the direction $(\ell_{\rm apex},b_{\rm apex})$ and the magnitude $(v_{\rm travel})$ of the disc motion by applying a Bayesian analysis of proper motions and radial velocities of hundreds of bright, distant stars in the smooth MW stellar halo: K Giants, Blue Horizontal Branch (BHB) stars and the MW satellites, each with individual 6D position-velocity information\cite{xue11,xue15,gaia18,mcconnachie20}.

\begin{figure*}[ht]
\centering
\includegraphics[width=\linewidth]{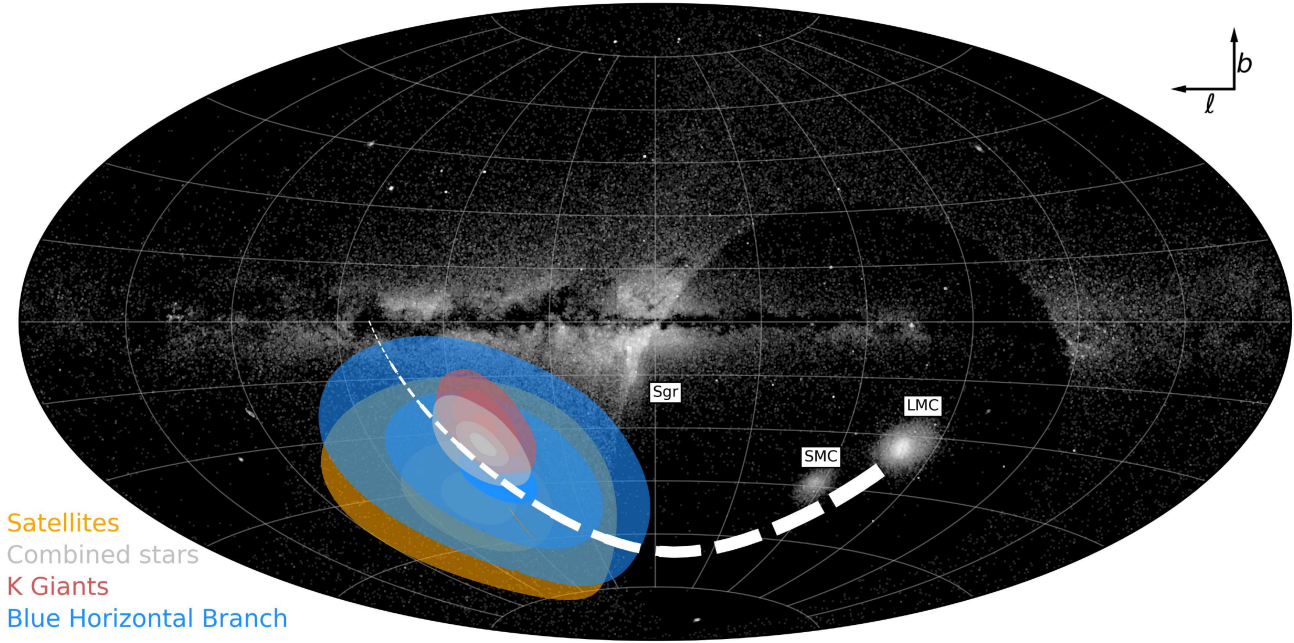}
\caption{{\bf Posteriors on the direction of the Milky Way disc motion relative to halo stars located at Galactocentric distances $r>40\kpc$, shown in Aitoff projection.} Shaded contours denote 67\%, 90\% and 95\% probabilities and are colour-coded according to each sample considered in this work. The black-and-white background image is the observed projected density of RR Lyrae stars identified in the union of the Pan-STARRS DR1 and Gaia DR2 catalogues \cite{sesar17,holl18,rimoldini18}. The LMC, as well as two less massive satellites, the Small Magellanic Cloud (SMC) and Sagittarius dwarf galaxy (Sgr), are labelled. Comparison with the plane of the LMC orbit derived from its proper motions (dashed white curve) indicates that the Milky Way disc is roughly moving towards a point along the past trajectory of the LMC across our Galaxy. The Pan-STARRS survey only covers the Northern sky, and thus the region around the LMC is only covered by Gaia and thus comparatively undersampled in RR Lyrae stars.}
\label{fig:apex}
\end{figure*}

Distant tracers have long dynamical times (see Supplementary Information) and are thus slow to react to the gravitational pull of the LMC\cite{petersen20a}. As the MW disc moves in response to the LMC, but the outer halo does not, a {\it reflex motion} arises in which the distant halo appears to move relative to the MW disc, when in fact it is the MW disc moving relative to the distant halo stars, an effect akin to the apparent motion of rain drops seen from a car driving through a downpour. Figure~\ref{fig:apex} shows the on-sky probability density function for the direction of the MW disc travel imprinted in the kinematics of each of the three halo tracers $(\ell_{\rm apex},b_{\rm apex})$, against a backdrop of RR Lyrae stars\cite{sesar17,holl18,rimoldini18}. All three tracers return consistent values at the 67\% confidence level (see Table~\ref{tab:posteriors}). Strikingly, the apex direction is not pointing towards the present-day location of the LMC on the sky, but rather along its historical trajectory across the Milky Way. As the LMC approached its current location with great speed, the MW disc was not able to keep up. As a result, the MW disc appears to be travelling in the direction of the LMC at some earlier point on its trajectory, as indicated by the plane of the LMC orbit (dashed curve). A similar effect is observed in the numerical models that follow the accretion of the LMC onto the MW (see Supplementary Information).

In Figure~\ref{fig:vtravel}, we show the probability density function of the magnitude of the disc motion $v_{\rm travel}$ derived from each of the studied tracers. The strongest constraint is provided by the K Giants, with $v_{\rm travel,~K~Giants}=35^{+5}_{-5} \kms$. Being the most numerous bright tracers in our data sample and having accurate proper motions out to large Galactocentric distances (see Extended Data Figure~\ref{fig:distances} and Supplementary Information), these stars currently provide the best kinematic targets of the outer-most halo. The full phase-space information also allows for removal of major substructure in the halo (see Extended Data Figure~\ref{fig:sgr}). BHBs and satellites return slightly less constraining measurements of the reflex motion, $v_{\rm travel,~BHB}=26^{+9}_{-9}\kms$ and $v_{\rm travel,~satellites}=55^{+22}_{-23}\kms$, respectively. All three data sets are consistent at the 67\% confidence level (see Table~\ref{tab:posteriors} and Extended Data Figure~\ref{fig:cornerplot}). Fitting the combined sample of BHBs and K Giants yields $v_{\rm travel,~combined}=32^{+4}_{-4}\kms$. The significant reflex motion and location of the apex is consistent with a massive LMC falling in for the first time. 
To test our Bayesian-fitting technique we use mock data sets of the stellar halo drawn from numerical models of the MW-LMC interaction (see Extended Data Figures~\ref{fig:mockcorner} and Supplementary Information). The magnitude of the reflex motion observed in the stellar halo of the Milky Way is consistent with numerical models where the infall mass of the LMC is greater than $10^{11}M_\odot$ (see Extended Data Figures~\ref{fig:apextraj}, \ref{fig:displace} and \ref{fig:time}), which is significantly larger than the mass enclosed within its luminous radius, $M_\lmc(<8.7\kpc)\simeq 0.17\times 10^{11}M_\odot$ [ref. \cite{vandermarel14}], suggesting that the LMC fell in surrounded by an extended dark matter halo.

\begin{figure}[ht]
\centering
\includegraphics[width=0.5\linewidth]{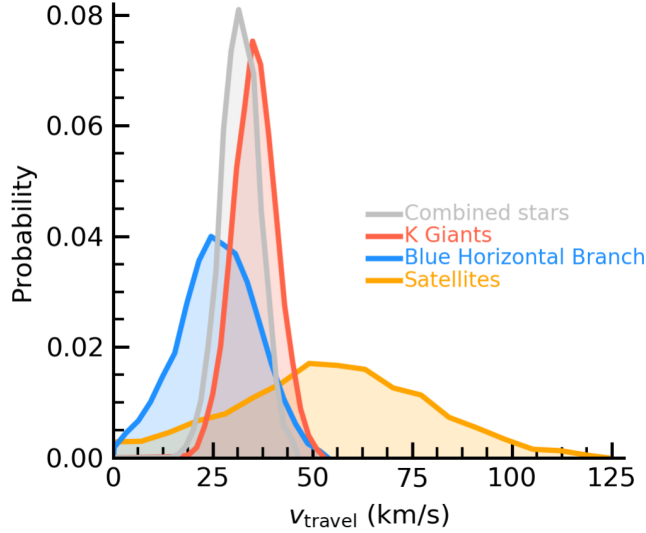}
\caption{{\bf Measured velocity of the Milky Way disc with respect to halo stars located at Galactocentric distances $r>40\kpc$ (red and blue curves), the combined sample (silver curve) and satellites (orange curve).} We find that the three distinct samples of halo tracers (K Giants, BHBs, and satellites) return consistent values within statistical uncertainties.}
\label{fig:vtravel}
\end{figure}

\begin{table}
\caption{{\bf Posteriors on model parameters.}\label{tab:posteriors}}
\begin{center}
\begin{tabular}{|l|c|c|c|c|c|c|c|}
\hline
Description & Parameter & Units & K Giants + BHBs & K Giants &  BHBs &  Satellites \\
\hline
Disc barycentre velocity & $v_{\rm travel}$ & $\kms$ & $32^{+ 4}_{- 4}$ &$35^{+ 5}_{- 5}$ &$26^{+ 9}_{- 9}$ &$55^{+22}_{-23}$ \\
Apex longitude &$\ell$ & deg & $56^{+ 9}_{- 9}$ &$53^{+ 9}_{- 9}$ &$59^{+25}_{-24}$ &$74^{+46}_{-34}$ \\
Apex latitude & $b$ & deg & $-34^{+10}_{- 9}$ &$-28^{+10}_{- 9}$ &$-42^{+19}_{-16}$ &$-49^{+20}_{-17}$ \\
Mean halo radial velocity & $\langle v_r\rangle$ & $\kms$ & $-18^{+ 4}_{- 4}$ &$-17^{+ 5}_{- 5}$ &$-22^{+ 8}_{- 8}$ &$-35^{+16}_{-16}$ \\
Mean halo azimuthal velocity & $\langle v_{\phi}\rangle $ & $\kms$ & $-20^{+ 5}_{- 4}$ &$-22^{+ 5}_{- 5}$ &$-9^{+ 9}_{- 9}$ &$-25^{+20}_{-20}$ \\
Mean halo polar velocity & $\langle v_{\theta}\rangle$ & $\kms$ & $14^{+ 5}_{- 5}$ &$18^{+ 6}_{- 6}$ &$-2^{+10}_{- 9}$ &$-19^{+25}_{-24}$ \\
Line-of-sight velocity hyperparameter & $\sigma_{h,{\rm los}}$ & $\kms$ & $93^{+ 2}_{- 2}$ &$94^{+ 3}_{- 3}$ &$91^{+ 3}_{- 4}$ &$92^{+13}_{-10}$ \\
Galactic longitude velocity hyperparameter & $\sigma_{h,\ell}$ & $\kms$ & $81^{+ 4}_{- 4}$ &$81^{+ 4}_{- 4}$ &$76^{+ 8}_{- 8}$ &$102^{+15}_{-12}$ \\
Galactic latitude velocity hyperparameter & $\sigma_{h,b}$ & $\kms$ & $76^{+ 3}_{- 3}$ &$71^{+ 3}_{- 3}$ &$96^{+ 8}_{- 7}$ &$108^{+16}_{-13}$ \\
\hline
\end{tabular}
\end{center}
\end{table}

Our results complicate theoretical studies of the MW stellar halo and firmly indicate that our Galaxy cannot be treated in dynamical equilibrium. In addition, Earth observers must correct for non-inertial effects introduced by the MW disc motion
when translating velocities measured in a heliocentric frame into a coordinate system whose origin is at the Galactic centre of mass. While corrections from the solar reflex are commonly applied, the reflex motion of the MW disc has so far been neglected in dynamical studies of distant objects. Failure to properly treat the LMC infall may bias measurements of various quantities, such as the total mass of the MW\cite{erkal20b}. The measurements presented here, constraining both the amplitude and direction of travel of the MW disc, will help to remove non-inertial perspective effects from the kinematics of the outer stellar halo and the satellite galaxy population. 

Previous studies have used similar data sets to those considered here to measure the rotation pattern and velocity dispersion profile of the stellar halo \cite{deason17,bird20}, but did not account for the reflex motion of the MW disc. A recent work\cite{erkal20b} finds a net positive $v_z$ motion in a sample of globular clusters and dwarf galaxies with 6D observations. As the LMC is expected to induce the disc to move downwards, one expects and upward motion in the outer halo. However, the magnitude and direction of the reflex motion signal has remained unconstrained to date. In particular, the localization of the apex signal is only enabled by combining full phase-space information of distant halo tracers (see Extended Data Figure~\ref{fig:apextraj}).

Our results conclusively show that the kinematics of halo stars at $r\ge 40\kpc$ are affected by the displacement of the MW disc from the Galactic barycentre. At smaller radii, the magnitude of reflex motion diminishes and the kinematic pattern becomes more dynamically convoluted than a simple dipole as the inner stellar halo reacts in response to the displacement of the disc potential\cite{petersen20a}. Tidal streams may also be affected by the dislodgement of the MW disc from the Galactic centre of mass\cite{vasiliev20}. Intriguingly, current constraints on the shape of the dark matter halo using tidal streams have so far reported contradictory values that cover all possible geometries: from spherical to oblate to prolate to even triaxial shapes\cite{johnston05,law10,bovy16,fardal19,malhan19}. Given that the resulting reflex motion varies across the sky (see Extended Data Figure~\ref{fig:modelsky} in Supplementary Information), and that tidal streams can wrap large regions of the Galaxy, this effect may be a significant factor to explain the discrepant measurements. 

Current state-of-the-art dynamical-modelling techniques often compute orbits in gravitational potentials centred at the MW disc in order to constrain the distribution of matter in our Galaxy. Our results indicate that time-varying MW-LMC potentials where both galaxies are allowed to deform due to their mutual gravitational attraction is a crucial theoretical step forward that must be taken. Once self-consistent MW-LMC potentials are developed, a number of questions open up: what was the structure of the stellar and dark matter haloes prior to the infall of the LMC? What is the exact trajectory of the LMC since it fell into the MW, and how did the dynamics of dark matter alter its path? Does the LMC lose its dark matter halo envelope to tides, and where does the stripped dark matter end up in our Galaxy? In this context, the detection of the reflex motion of the MW disc calls for developing more sophisticated, self-consistent models that capture non-equilibrium features in the stellar halo. For example, the LMC is predicted to induce strong perturbations on the spatial configuration and kinematics of stellar tracers, such as the formation of a pronounced `wake' in the smooth stellar halo\cite{garavitocamargo19}, and to perturb tidal streams out of their orbital planes\cite{erkal19}.

Astronomers must also turn their attention to the dynamics of the Local Group, a larger volume of galaxies that includes our nearest massive neighbor, Messier 31 (M31, the Andromeda galaxy). The quadrant that hosts M31 ($\ell<180^\circ,b<0^\circ$) is one of the regions of the sky most affected by reflex motion, a finding that suggests that previous studies treating the orbit of M31 will need to be revised in light of the localized direction of the MW disc displacement with respect to the Galaxy barycentre.

 Our understanding of the MW stellar halo will be revolutionized in the next decade.
 Upcoming data releases from the Gaia mission will provide robust proper-motion measurements from the solar volume out to the most distant regions of the Galaxy. When coupled with ongoing and future radial velocity surveys (LAMOST\cite{deng12}, 4MOST\cite{dejong19}) and new instruments (MOONS\cite{cirasuolo14}), the number of halo stars with accurate 6D phase-space measurements could be potentially increased by a factor $\sim 100$ (see Supplementary Information). Such data sets will allow a precise measurement of the radial variation of reflex motion, and reveal the infall trajectory of the LMC as well as the location of its tidal debris, which in turn will constrain the dark matter distribution in both galaxies with unprecedented accuracy.

\renewcommand{\figurename}{Extended Data Fig.}
\setcounter{figure}{0}

\section*{Methods}

\subsection*{Bayesian inference method}
In this section we describe the fundamentals of a Bayesian method to search for signatures of reflex motion in the outer stellar halo. We analyse stars with full phase-space information, i.e. heliocentric distances ($D$), Galactocentric coordinates $(l,b)$, line-of-sight velocities ($v_{\rm los}$), and proper motions, $(\mu_\ell,\mu_b)$. The technique is applied to observational data sets as well as to mock catalogues of halo stars created from $n$-body realizations of the LMC-MW system (see Supplementary Information).

Our models contain nine free parameters that are fitted simultaneously to the data. These are the magnitude of the disc velocity, $v_{\rm travel}$, its apex direction in Galactocentric coordinates $(\ell_{\rm apex},b_{\rm apex})$, the mean velocity of halo stars in spherical coordinates with origin at the Galactic centre $\langle v_r\rangle$, $\langle v_\phi\rangle$ and $\langle v_\theta\rangle$, plus three hyperparameters $\sigma_{\rm h,los}$, $\sigma_{h,\ell}$ and $\sigma_{h,b}$, with units of $\kms$, which are added in quadrature to the observational measurement variance in order to minimize covariance with the rest of model parameters. Hyperparameters provide a useful tool to assign weights to data sets beyond those derived from statistical errors\cite{hobson02}. Their chief role is to minimize biases arising from (unknown) model imperfections by increasing the uncertainty associated with the posterior distributions. These statistical objects effectively act as `nuisance' parameters with no direct physical meaning, which are fitted to the data to be subsequently marginalized over the posterior distributions on the quantities of interest. 

 Our analysis allows for non-zero mean velocities in each degree of freedom, $\langle v_r\rangle$, $\langle v_\phi\rangle$ and $\langle v_\theta\rangle$, as well as a reflex motion of the Milky Way disc owing to the gravitational pull of the LMC, $\Delta \vec{ v}_{\rm travel}$. The motion of the Milky Way disc with respect to the stellar halo is modelled as a simple dipole. Comparison against live $n$-body models shows that this analytical model describes the reflex motion of halo stars at large Galactocentric distances remarkably well\cite{petersen20a}. To this aim, we rotate the Galactocentric reference system in such a way that the new $z$-axis is aligned with the disc motion, which points in the direction $(\ell_{\rm apex},b_{\rm apex})$. The spherical azimuthal and polar coordinates of any given star are transformed to the rotated frame as $(\phi,\theta)\to (\phi_1,\phi_2)$. In this frame, the dipole $\Delta \vec{ v}_{\rm travel}=(\Delta v_r,\Delta v_{\phi_1},\Delta v_{\phi_2})$ can be simply computed as

\begin{align}
    \Delta v_r&=-v_{\rm travel}\,\cos(\phi_2)\\ \nonumber
    \Delta v_{\phi_2}&=+v_{\rm travel}\,\sin(\phi_2) \\ \nonumber
    \Delta v_{\phi_1}&=0.
\end{align}
The rotation of the Galactocentric axes is done using an Euler-XYZ transformation with angles $(\phi,\theta,\psi)_{\rm rot}=(l_{\rm apex},\pi/2-b_{\rm apex},0)$.
Conversion between heliocentric and Galactocentric frames is done by adopting a right-handed Cartesian coordinate system where the centre of the MW is at $(x,y,z)=(0,0,0)$, and the sun is at $\vec{r}_{\odot\to\mw}= (-8.3,0.,0.02)$ kpc (refs. \cite{gravity19,bennett19}).  In these coordinates the sun moves with a velocity $\vec{v}_{\odot\to\mw}=(11.1,244.24,7.25) \kms$ (refs. \cite{schonrich10,mcmillan17}). Changes of up to a few per cent to these quantities do not affect our analysis\cite{drimmel18}; however, $v_{\rm travel}$ is modestly dependent on the choice of the rotation velocity at the solar circle such that increasing the rotation velocity at the solar circle decreases the magnitude of the reflex motion, and vice versa. 

In a heliocentric coordinate system, the average velocity vector of the stellar halo can be calculated from the mean velocity components fitted in our analysis and the reflex motion of the disc as

\begin{align}\label{eq:averv}
    \langle \vec{v} \rangle=\Delta \vec{v}_{\rm travel}+\langle \vec{v}_r\rangle+\langle \vec{v}_\phi\rangle+\langle \vec{v}_\theta\rangle -\vec{v}_{\odot\to\mw}.
\end{align}
 The observed components of the vector~(\ref{eq:averv}) correspond to the following projections

\begin{align}
    \langle v_{\rm los} \rangle&= \langle \vec{v} \rangle\cdot \hat u_{\rm los} \\ \nonumber
    \langle \mu_l \rangle&= \frac{\langle \vec{v} \rangle\cdot \hat u_{\rm \ell}}{k D} \\\nonumber
    \langle \mu_b\rangle&= \frac{\langle \vec{v} \rangle\cdot \hat u_{\rm b}}{kD},
\end{align}
where $\hat u_{\rm los}$, $\hat u_{\ell}$ and $\hat u_{\rm b}$ are unit vectors aligned with the line-of-sight direction and the Galactocentric coordinates $(\ell,b)$, respectively. We use the conversion from velocity to proper motion

\begin{align}
v_i=k D \mu_i ,
\end{align}
with a sub-index $i=\ell,b$ and a geometrical factor $k=4.74057 \kms\kpc^{-1}({\rm milliarcsec/\yr})^{-1}$. Here, $D$ is the heliocentric distance measured in $\kpc$.

To derive posteriors on the model parameters we adopt a Gaussian likelihood 
\begin{align}\label{eq:prob}
    \mathcal{L}(\{D_i,\ell_i,b_i,v_{\rm los,i},\mu_{\ell,i},\mu_{b,i}\}^{N_{\rm sample}}_{i=1}|\vec{S})=\prod_{i=1}^{N_{\rm sample}} p_{\rm los}(v_{\rm los})p_{\rm pm}(\mu_\ell,\mu_{b});
\end{align}
where $\vec{S}=(v_{\rm travel},\ell_{\rm apex},b_{\rm apex},\langle v_r\rangle,\langle v_\phi\rangle,\langle v_\theta\rangle,\sigma_{\rm h, los},\sigma_{{\rm h},\ell},\sigma_{\rm h,b})$ is a vector that comprises the model parameters. $p_{\rm los}$ is the 1D normal probability functions associated with line-of-sight velocities

\begin{align}
 p_{\rm los}(v_{\rm los})&=\frac{1}{\sqrt{2\pi \sigma_{\rm los}^2}}\exp\bigg[-\frac{(v_{\rm los}-\langle v_{\rm los}\rangle)^2}{2\sigma_{\rm los}^2}\bigg]; 
\end{align}
with a 1D variance along the line-of-sight direction

\begin{align}\label{eq:sigma_rad}
\sigma^2_{\rm los}&=\epsilon_{\rm los}^2 + \sigma_{\rm h,los}^2,
\end{align}
where $\epsilon_{\rm los}$ is the error associated to the line-of-sight velocity, plus the additional ‘freedom’ provided by the hyperparameter $\sigma_{\rm h, los}$. 

The likelihood associated with proper motions is modelled as a bivariate Gaussian,

\begin{align}\label{eq:probpm}
 p_{\rm pm}(\mu_\ell,\mu_{b})&=\frac{1}{2\pi \sqrt{\det(C) }}\exp\bigg[-(\chi-\langle{\chi}\rangle)^TC^{-1}(\chi-\langle{\chi}\rangle)\bigg];
\end{align}
where $\chi=(\mu_\ell,\mu_b)$ is the data vector, and $\langle\chi\rangle=(\langle\mu_\ell\rangle,\langle\mu_b\rangle)$ is the vector containing the systemic proper motions. The covariance matrix, $C$, contains the correlation between the proper motion errors

\begin{align}\label{eq:cov}
C=
\begin{bmatrix}
\epsilon'^2_{\ell} & \epsilon_{\ell}\epsilon_b\rho\\
\epsilon_{\ell}\epsilon_b\rho & \epsilon'^2_{b}
\end{bmatrix} 
\end{align}
where $\rho$ and $\epsilon_{\ell,b}$ are correlation coefficients and measurement uncertainties, respectively.
The 2D variances in proper motion space ($\epsilon'_{\ell,b}$) can be estimated as

\begin{align}\label{eq:sigma_prop}
\epsilon'^2_{\ell}&=\epsilon_{\ell}^2 + \epsilon_D^2\bigg|\frac{\d \mu_\ell}{\d D}\bigg|_D^2+\frac{\sigma_{{\rm h},\ell}^2}{(kD)^2}=\epsilon_{\ell}^2 + \epsilon_D^2\frac{|\mu_\ell|^2}{D^2}+\frac{\sigma_{{\rm h},\ell}^2}{(kD)^2}\\ \nonumber
\epsilon'^2_{b}&=\epsilon_{b}^2 + \epsilon_D^2\bigg|\frac{\d \mu_b}{\d D}\bigg|_D^2+\frac{\sigma_{{\rm h},b}^2}{(kD)^2}=\epsilon_{b}^2 + \epsilon_D^2\frac{|\mu_b|^2}{D^2} +\frac{\sigma_{{\rm h},b}^2}{(kD)^2},
\end{align}
which include proper motion ($\epsilon_{\ell,b}$) as well as distance ($\epsilon_D$) errors\cite{ma13}. In practice, we find that the the variances $\epsilon'^2_{\ell}$ and $\epsilon'^2_{b}$ are typically dominated by the nuisance parameters $\sigma^2_{{\rm h},\ell}$ and $\sigma^2_{{\rm h},b}$, respectively. These statistical parameters absorb random motions in the stellar halo, which in general are much higher than the velocities associated with measurement errors. 

We adopt flat priors for $v_{\rm travel},\ell_{\rm apex},\cos(b_{\rm apex}),\langle v_r\rangle,\langle v_\phi\rangle$ and $\langle v_\theta\rangle$, and Jeffreys priors for $\sigma_{\rm h,los}, \sigma_{{\rm h},\ell}$ and $\sigma_{{\rm h},b}$, with ranges that include reasonable parameter values. 

\begin{figure*}[ht]
\centering
\includegraphics[width=0.75\linewidth]{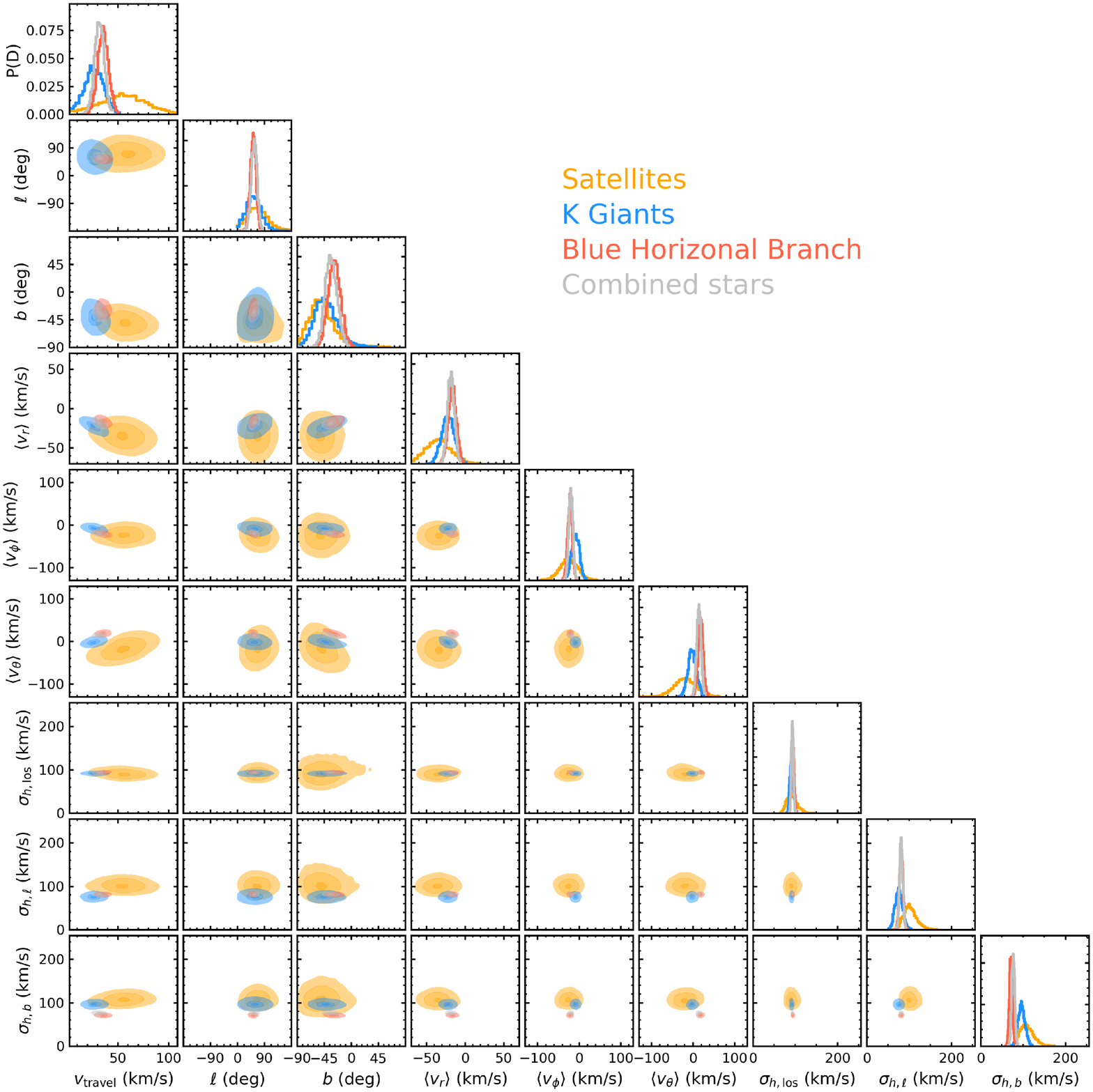}
\caption{{\bf Corner plot of covariance in the nine fit parameters for the Blue Horizontal Branch (blue), K Giant (red), satellite (orange) and combined K Giant+Blue Horizontal Branch  samples (silver).} The fit parameters are disc velocity, $v_{\rm travel}$; Galactocentric longitude apex $\ell_{\rm apex}$; Galactocentric latitude apex $b_{\rm apex}$; the mean velocity of halo stars in spherical coordinates $\langle v_r\rangle$, $\langle v_\phi\rangle$ and $\langle v_\theta\rangle$; the three hyperparameters $\sigma_{\rm h,los}$, $\sigma_{h,\ell}$ and $\sigma_{h,b}$. A full description of the model may be found in the Supplementary Information. \label{fig:cornerplot}}
\end{figure*}

See Extended Data Figure~\ref{fig:cornerplot} for full covariances between the nine fit parameters.

\subsection*{Inference code}
We apply a nested-sampling technique\cite{skilling04} to calculate posterior distributions for our parameters and the evidence of the model. Here, we use the code {\sc MultiNest}, a Bayesian inference tool which also produces posterior samplings and returns error estimates of the evidence\cite{feroz08,feroz09}.

\begin{figure}[ht]
\centering
\includegraphics[width=0.6\linewidth]{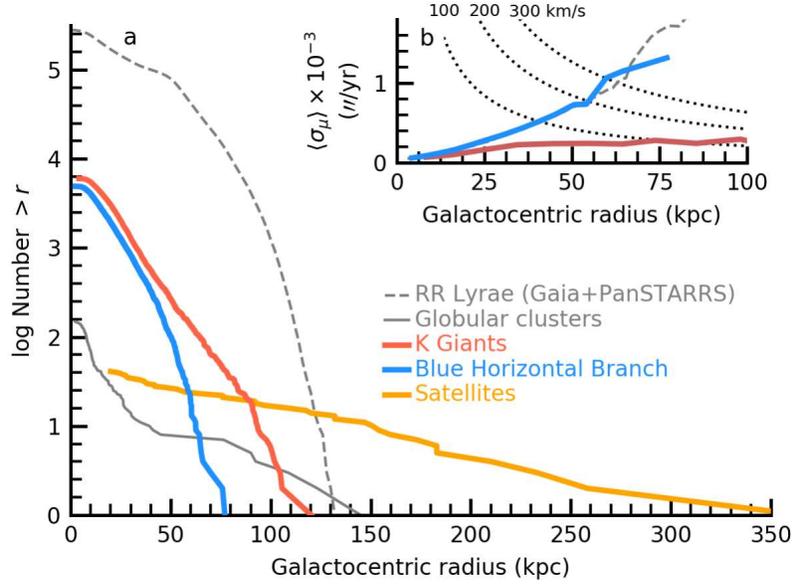}
\caption{{\bf Tracer population properties as a function of distance.} Main panel: logarithm of number of sources larger than some Galactocentric radius, $r$, as a function of Galactocentric radius $r$ in kiloparsecs. We show the three data sets analyzed in this paper in color (red: K Giants, blue: Blue Horizontal Branch, orange: satellites). We show two explored, but unused, data sets in gray (solid gray: globular clusters, dashed gray: RR Lyrae stars). Inset: the mean proper motion error in milliarcseconds per year for the stellar sources (K Giants, Blue Horizontal Branch, RR Lyrae), as a function of Galactocentric radius. As the brightest sample, K Giants have the smallest uncertainty. We mark 100, 200, and 300 $\kms$, the approximate range of halo velocities, as a dotted black curves. \label{fig:distances}}
\end{figure}

\begin{figure}[ht]
\centering
\includegraphics[width=0.7\linewidth]{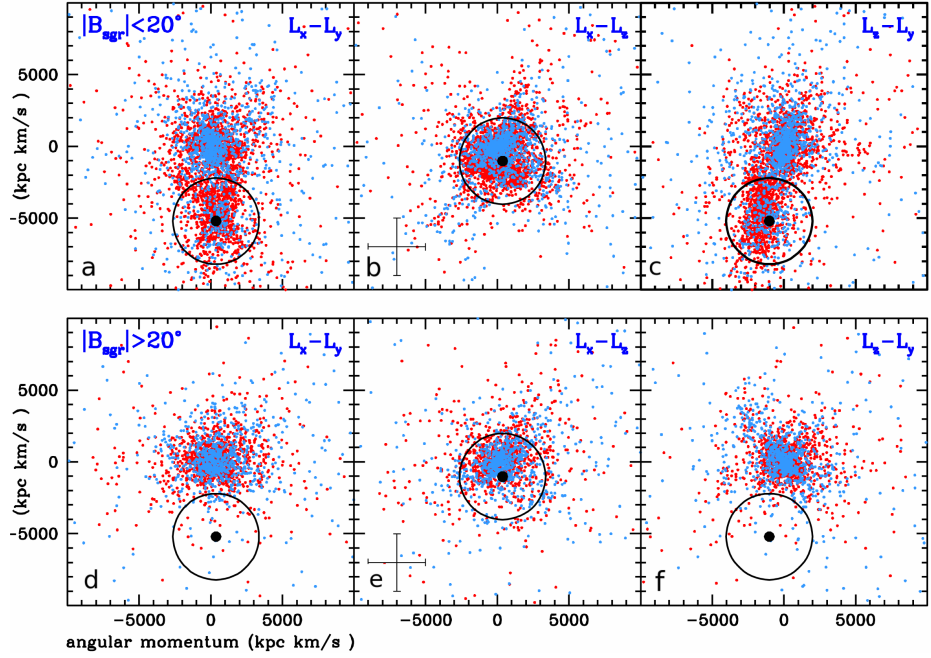}
\caption{{\bf Angular momentum components of halo stars at Galactocentric distances $r>20\kpc$.} Blue and red dots denote BHBs and K Giants, respectively. The angular momentum of the Sagittarius dwarf is marked with a black dot at $\vec L_{\rm sgr}/(\kpc\kms)=(605, -4515, -1267)$. For reference, circles mark angular momentum difference $|\vec{L}-\vec{L}_{\rm sgr}|=3000\,\kpc\kms$. Upper and lower panels correspond to stars at $|B_{\rm sgr}|<20^\circ$ and $|B_{\rm sgr}|>20^\circ$ off the orbtital plane of the Sagittarius stream, respectively. We show the mean 1$\sigma$ error bar at $r>40 \kpc$ in the lower left of the centre column. \label{fig:sgr}}
\end{figure}

\begin{figure}[ht]
\centering
\includegraphics[width=0.75\linewidth]{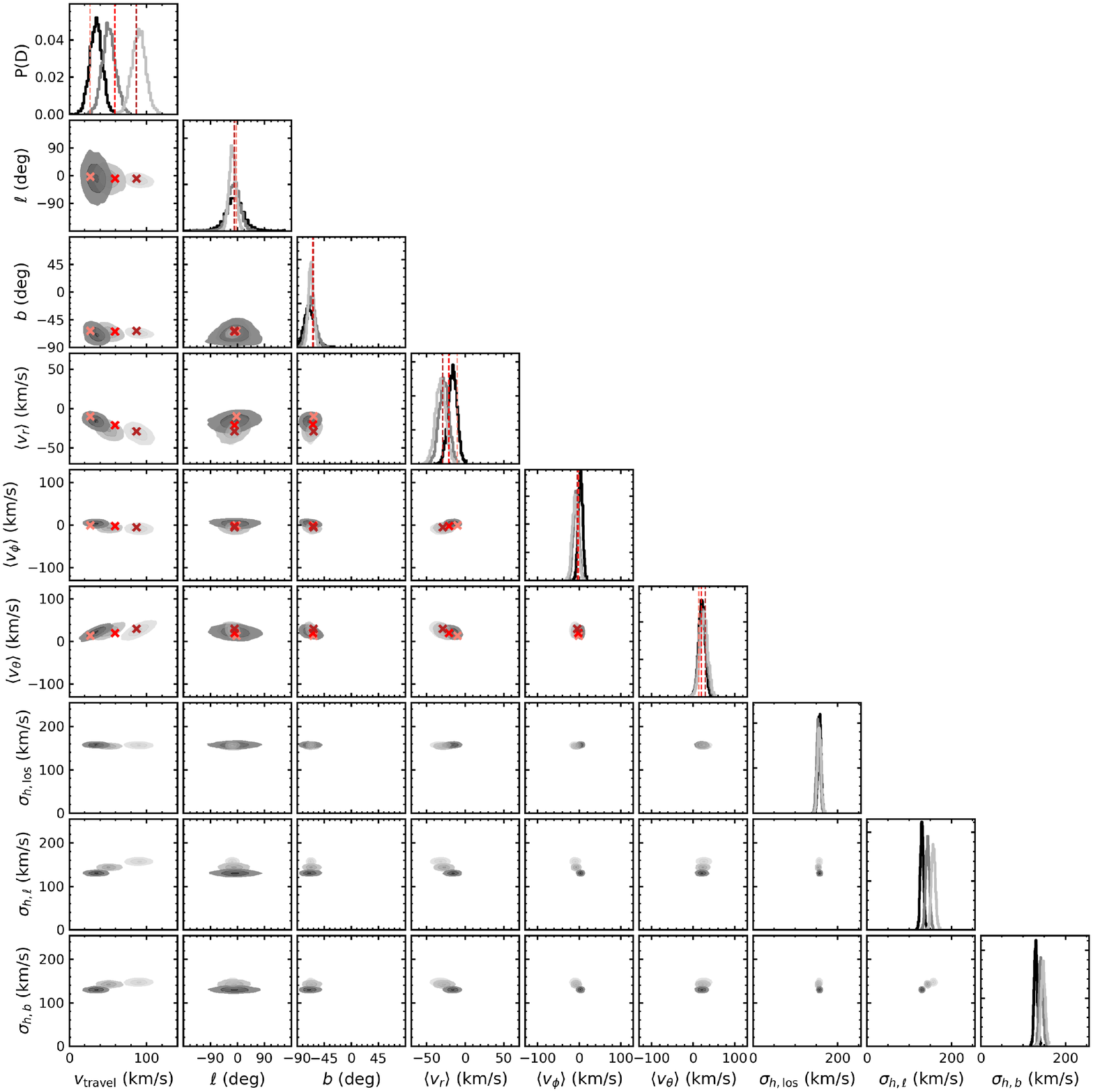}
\caption{{\bf Corner plot for mock data sets of K Giant stars at $r>40\kpc$ drawn from live $n$-body simulations of the MW where the LMC falls in with a mass $M_\lmc/(10^{11}M_\odot)=1,2$ and 3.} See text for mock data details. All parameters are well constrained and show a relatively minor covariance. Note that the bounds on the apex direction and the magnitude of the reflex motion improve in proportion to the LMC mass. Values derived directly from the simulation are shown as dashed vertical lines (on histograms), or coloured `x' markers on contour plots. \label{fig:mockcorner}}
\end{figure}

\begin{figure}[ht]
\centering
\includegraphics[width=0.85\linewidth]{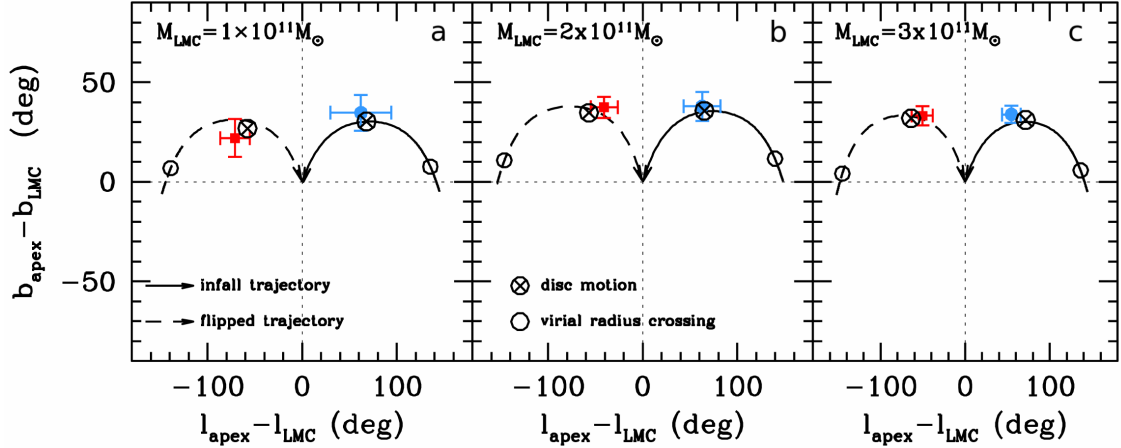}
\caption{{\bf Apex direction measured from mock samples of K Giant stars located at $r>40 \kpc$ in the SDSS footprint.} We use live Galaxy models that experience the infall of LMC-like galaxies with masses $M_{\rm LMC}/10^{11}M_\odot=1,2$ and 3 (left, middle and right panels, respectively). Solid lines denote the infall trajectory of the LMC derived from backwards orbit integration of HST proper motions, while dashed lines show models where the LMC trajectory has been flipped. Red (blue) symbols denote the measurement for mock samples with an LMC-like (flipped) trajectory. Uncertainties are the standard deviation derived from the posteriors of the fit locations. Open circles show the point on the trajectory where the LMC crossed the virial radius of the Galaxy. Note that within statistical uncertainties the apex direction derived from the kinematics of distant stellar halo particles points towards the direction where the MW disc is currently moving (crossed circles). The disc component is currently travelling to an earlier point on the LMC trajectory. 
\label{fig:apextraj}}
\end{figure}

\begin{figure}[ht]
\centering
\includegraphics[width=0.5\linewidth]{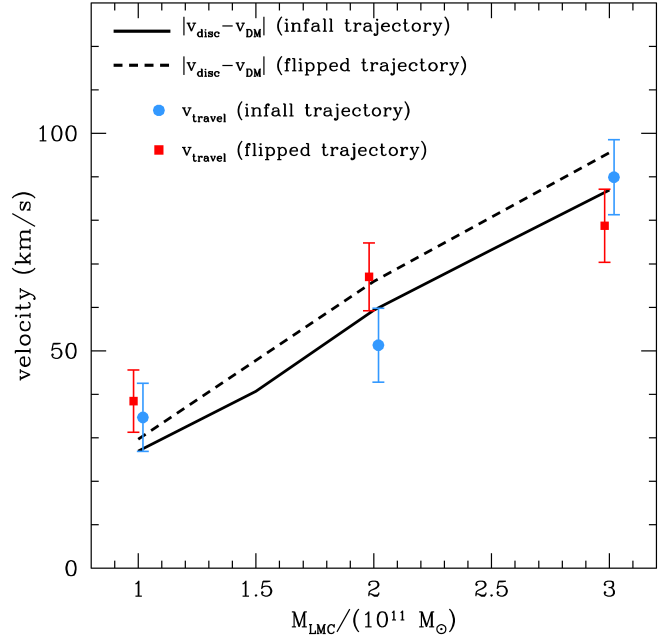}
\caption{{\bf Velocity of the Galactic disc ($v_{\rm travel}$) inferred from the kinematics of stellar tracers located at $r>40 \kpc$ as a function of LMC mass.} The solid (dashed) lines show the true speed of the disc barycentre relative to dark matter particles within a radial range $40<r/\kpc<150$ on the infall (flipped) trajectories in the mock LMC models. The blue (red) symbols show the value of $v_{\rm travel}$ measured from the mock models in the infall (flipped) trajectory case. \label{fig:displace}}
\end{figure}

\begin{figure}[ht]
\centering
\includegraphics[width=0.5\linewidth]{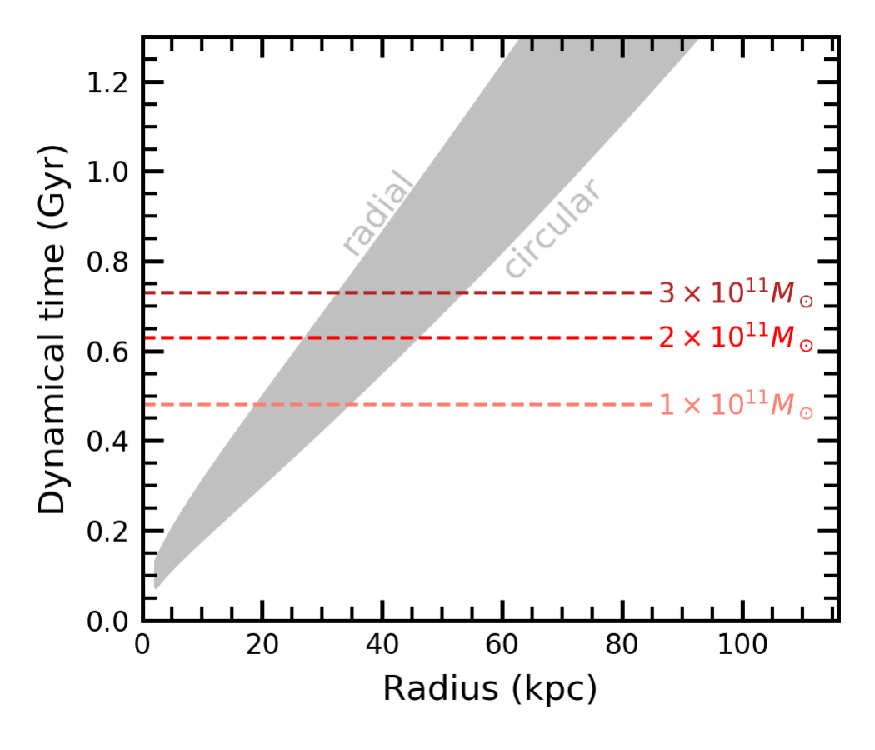}
\caption{{\bf Dynamical time of particles in the model MW-LMC systems as a function of Galactocentric radius.} For ease of reference, horizontal dashed lines show the lookback time for the onset of disc motion. Halo particles with a dynamical time above the horizontal lines react impulsively to the LMC infall. At a fixed radius, radial and circular orbits provide the shortest and longest periods, respectively. As a function of radius, a star's period will fall in the grey shaded region. \label{fig:time}}
\end{figure}

\begin{figure*}[ht]
\centering
\includegraphics[width=0.9\linewidth]{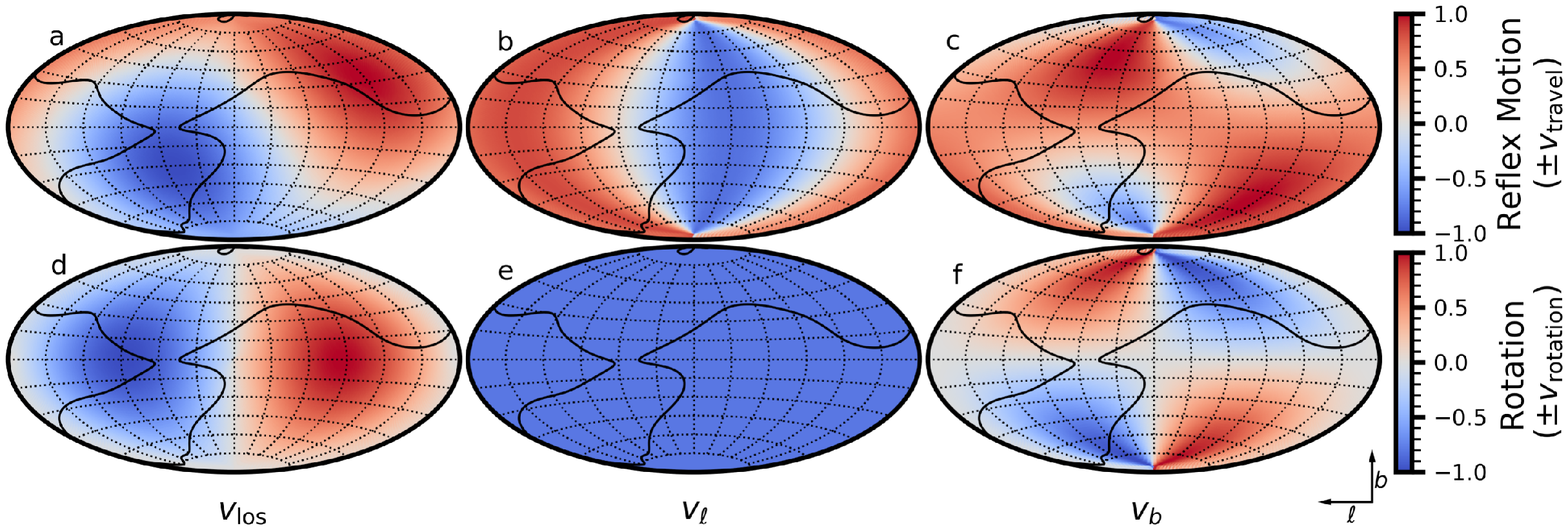}
\caption{{\bf Comparison of the appearance of reflex motion versus rotation.} Upper row: our reflex motion model, projected onto the $(\ell,b)$ plane in $v_{\rm los},v_\ell,v_b$. Velocities have been normalized such that the extreme values are $v_{\rm travel}$. Lower row: a rotation model, with magnitude $v_\phi$, projected onto the ($\ell,b$) plane in $v_{\rm los},v_\ell,v_b$. The smoothed SEGUE footprint is outlined in black. Comparing the upper and lower rows shows that the line-of-sight and $b$ velocities are similar, supporting the need for 6D data and all-sky coverage. \label{fig:modelsky}}
\end{figure*}

\subsection*{Data sets}

We use available data sets of two luminous stellar tracers at $r>40$ kpc: Blue Horizontal Branch (BHB) stars \cite{xue08,xue11}, and K Giants \cite{xue14,xue15} identified using SDSS/SEGUE photometry and spectroscopy \cite{yanny09}. We accept the positions, distances, and uncertainties in the published tables. We apply a color-cleaning technique to reduce Blue Straggler contamination in the BHB sample, eliminating stars that fall in the region of color space suggested in ref. \cite{lancaster19}. We cross-match both the BHB and K~Giant catalogues at 1$^{\prime\prime}$ tolerance with Gaia DR2 \cite{gaia16, gaia18} to obtain proper motions in right ascension and declination. We filter the samples by {\tt ruwe}, the reduced unit weighted error -- a collected measure of the data quality -- accepting only stars with {\tt ruwe}$<1.4$ \cite{lindegren18}. We convert the observed proper motions to Galactocentric coordinates, including proper motion uncertainty covariance.

While Gaia DR2 has known systematic errors in proper motion for bright stars $(G<12)$, the systematics for faint stars do not appear to show any large-scale structure that would bias our results and can be assumed to be in a non-rotating frame\cite{lindegren18}. We accept the reported mean and root-mean-square all-sky uncertainties (in milliarsecond yr$^{-1}$: $\langle\mu_{\rm RA}\rangle=0.;\sigma_{\rm RA}=0.039;\langle\mu_{\rm Dec}\rangle=0.011;\sigma_{\rm Dec}=0.037$) and emphasize that the root-mean-square uncertainties are an order of magnitude smaller than the typical errors on individual sources. 

Although the catalogues do not distinguish between disc and halo stars, we do not expect any disc stars at distances $r>40\kpc$. Independently of published distances, we have also checked that BHB and K Giant stars in our catalogue have unresolved Gaia parallaxes, which confirms their location beyond the disc volume. We also exclude stars with latitudes $|b|<10^\circ$ to minimize contamination from faint disc stars misidentified as BHBs or K Giants in the published catalogues.

The resulting catalogue still contains stars whose distances may be highly overestimated. These objects stand out as stars with very large velocities.
 In order to remove contaminants we have perform iterative sigma clipping follow a standard procedure: for all stars in the sample, we compute the difference between model prediction and measurement $\langle (data-model)/\sigma\rangle_i$, where brackets denote avarages over posterior distributions, and the subindex $i$ runs over the 3 velocity dimensions: radial velocities and the two proper motion components.
Here, $\sigma$ is the fitted dispersion, which combines individual measurement errors and the hyperparameters (eqs~\ref{eq:sigma_rad} and~\ref{eq:sigma_prop}). Stars with $|\langle (data-model)/\sigma\rangle_i|>3$ in any of those dimensions are considered {\it outliers}. After removing them, we re-fit the new sample. The above steps are iterated until convergence is achieved.

Give that the BHB and K Giant samples are drawn from SEGUE\cite{yanny09}, the sky coverage is not complete. We test the impact of incomplete sky coverage on our Bayesian inference method using mock catalogues drawn from live $n$-body simulations within the SEGUE footprint (see Extended Data Figures~\ref{fig:apextraj} and \ref{fig:displace} as well as Supplementary Information). The results indicate that the limited sky area does not introduce any meaningful bias. Additionally, we measure the magnitude of proper motions at $r>40$ kpc from the K Giant sample, finding that the proper motions are less than 1 milliarcsecond per year. This provides an estimate of the precision in proper motions required to studying an ensemble of stars in the MW halo at $r>40\kpc$. The canonical velocity of a star in the MW halo, $v=250 \kms$, corresponds to a proper motions of $\mu\approx1$ milliarcsecond per year at $50~\kpc$ (Extended Data Figure~\ref{fig:distances}).

Since we model the stellar halo as a smooth object, the presence of kinematic substructures, such as tidal streams, can potentially bias our results. The Sagittarius stream is the largest known substructure in the stellar halo of our Galaxy\cite{niedersteostholt10}. We exclude Sagittarius stream stars from our sample on the basis of their angular momentum coordinates  (see Supplementary Information for additional details). In short, wirst we identify stars located in the plane of the Sagittarius stream by rotating our reference frame such that the polar angle of stream stars measured from the Galactic centre is located at $B_{\rm sgr}=0$ on average \cite{majewski03}. Subsequently, we identify stars in the orbital plane of the stream by choosing those that fall within $|B_{\rm sgr}|<20^\circ$, and compute their angular momentum components, $\vec{L}=(L_x,L_y,L_z)$. Next, we draw a sphere with radius $3000\kpc\kms$ around the location of the Sagittarius dwarf $\vec{L}_{\rm sgr}=(605,-4515,-1267) \kpc\kms$. Stars that fall in this region are tagged as Sagittarius members and removed from our sample (Extended Data Figure~\ref{fig:sgr}). We have checked that the result does not strongly depend on the size of the sphere. We do not find that any of the stars in our sample belong to known globular clusters or satellite galaxies, apart from Sagittarius.  

We also use the positions, distances, radial velocities, and proper motions of MW satellites compiled from literature sources\cite{fritz18,riley19,pace19,fritz19,torrealba19,simon20,mcconnachie20}. To identify possible LMC satellites we follow a similar procedure as we did above to identify Sagittarius stream members. First, we locate the LMC and SMC in angular momentum space at $\vec{L}_\lmc=(-16044, 1236, -1520)\kpc\kms$ and $\vec{L}_\smc=(-15169, -3155, -2827)\kpc\kms$ and define the separation $|\Delta \vec{L}_{\rm MC}|\equiv |\vec{L}_\smc-\vec{L}_\lmc|=4664 \kpc\kms$ as a threshold in angular momentum space. Subsequently, we compute the angular momentum of the rest of MW satellites and rank-order them according to their separation with respect to the LMC, $|\vec{\Delta L}|=|\vec{L}-\vec{L}_\lmc|$. The first two galaxies in the ranking (Horologium~I and Carina~I) have an angular momentum separation $|\vec{\Delta L}|<|\Delta \vec{L}_{\rm MC}|$, which establishes a probable association. Consequently, Horologium~I and Carina~I are removed from the sample. A third galaxy, Phoenix~II, has been recently identified as a possible satellite of the LMC\cite{fritz19}, and indeed appears as the third galaxy in the ranking with a $|\vec{\Delta L}|\approx |\vec{\Delta L}_{\rm MC}|$. Accordingly, Phoenix~II is also excluded from the analysis. Following suggestions from the literature of satellites that may be strongly affect by the LMC, we also remove Carina~II, Carina~III, Hydrus~I, Hydra~II, Draco~II, Reticulum~II, Sculptor, Tucana~III, and Segue~1\cite{kallivayalil18,erkal19,patel20}.

Our final data sets consist of $N_{\rm K Giants}=543$, $N_{\rm BHB}=292$ and $N_{\rm satellites}=33$. Other stellar tracers, such as RR Lyrae stars, do not have accurate 6D information, including those with available radial velocities \cite{cohen17,sesar17,holl18,rimoldini18}. Distances for RR Lyrae stars are estimated using an absolute magnitude calibration tuned for Gaia\cite{iorio19}. Another common halo tracer, the MW globular cluster system, has available 6 dimensional information\cite{vasiliev19}, but only a handful of clusters are currently located at $r>40$ kpc (see Supplementary Information). We collect these data sets in the hope that future data-taking projects will expand the usable 6D stellar sample with an all-sky coverage. In particular, an all-sky sample of RR Lyrae with spectroscopic follow-up would provide a particularly rich data set, for which Gaia DR3 may return proper motions usable for our analysis beyond $\sim 40\kpc$ in the near future.

\pagebreak
\section*{Supplementary Information} 

\subsection*{Model parameters}

The Bayesian inference includes nine parameters: $v_{\rm travel}$, $\ell_{\rm apex}$, $b_{\rm apex}$, $\langle v_r\rangle$, $\langle v_\phi\rangle$, $\langle v_\theta\rangle$, $\sigma_{\rm h,los}$, $\sigma_{h,\ell}$, $\sigma_{h,b}$. The first three parameters, $v_{\rm travel}$, $\ell_{\rm apex}$, $b_{\rm apex}$, define the magnitude and direction of the disc motion with respect to the stellar halo. The mean velocities $\langle v_r\rangle$, $\langle v_\phi\rangle$, $\langle v_\theta\rangle$, account for bulk motions within the stellar halo in the radial, azimuthal and polar directions. Coherent motions are observed in tests against mock data drawn from live $n$-body models where the MW halo is allowed to react to the LMC infall (see below). For example, numerical models show a slight contraction of the stellar halo, $\langle v_r\rangle<0$, as a response to the LMC gravitational pull. In addition, a net tangential motion can originate from different sources, such as an inherent rotational component in the tracer population, as well as the presence of kinematic substructure in the stellar halo. 
The last three parameters, $\sigma_{\rm h,los}$, $\sigma_{h,\ell}$, $\sigma_{h,b}$ correspond to independent hyperparameters in each observational degree of freedom. These hyperparameters account for statistical scatter with physical origin, such as the velocity dispersion of the halo, as well as systematic errors introduced by imperfections in the theoretical model. For example, deviations from the dipole signal owing to higher-order spherical harmonics will be partially absorbed by the hyperparameters.

In Extended Data Table~\ref{tab:posteriors}, we list the median and 67\% confidence intervals of the posterior distributions on each model parameter. In Extended Data Figure~\ref{fig:cornerplot}, we show the full corner plot for all nine parameters included in the Bayesian inference. All parameters are reasonably well constrained, with relatively small degeneracies among them. 

The BHB and K Giant samples have reported metallicities. We investigated the natural (equal-number) divide between metal-poor and metal-rich stars ([Fe/H]$\approx -1.8$). We find that the metal-poor component is much less constrained owing to the larger errors in radial velocity measurements, which are on average twice larger than for metal-rich stars. As the LMC infall is a recent event, we do not expect any difference in reflex motion between the metal-poor and metal-rich stars in the halo, all of which should be uniformly perturbed by the reflex motion.

The upper panels in Extended Data Figure~\ref{fig:modelsky} show the dipole model returned by the median values of the parameters $\ell_{\rm apex},b_{\rm apex}$ given in Extended Data Table~\ref{tab:posteriors}, with solid lines marking the defined smooth SEGUE footprint. For comparison, lower panels plot the signal expected from a halo rotating around the $z$-axis. Extended Data Figure ~\ref{fig:modelsky} reveals key features of our results. Note first the striking similarity between the dipole and rotation models in the line-of-sight and polar coordinates. In those directions, incomplete sky coverage severely limits the area where the two models can be distinguished. The second point to notice is key relevance of the azimuthal component of the velocity vector, $v_\phi$. Here, rotation implies a constant value of $v_\phi$ across the sky, whereas our best-fitting dipole model shows a clear variation along the Galactocentric longitude, $\ell$. That SEGUE covers the full range of longitudes in the Northern hemisphere allows a clear-cut distinction between rotation and translation in the stellar halo when modelling the kinematics of halo stars with accurate proper motions. Extended Data Figure~\ref{fig:modelsky} also indicates that Gaia EDR3 data, which is expected to halve current proper motion errors, will significantly improve the constraints on the disc reflex motion presented in this work.

\subsection*{Sagittarius stream and substructure}
We model the stellar halo as a smooth object devoid of substructure. This assumption is clearly at odds with the presence of the Sagittarius stream, which comprises a significant fraction of the stellar halo\cite{niedersteostholt10}. Here, we apply cuts in angular momentum  space in order to remove stream contaminants from our data sample. We do not adopt any existing theoretical model for the Sagittarius dwarf to identify stream members because all existing models have been shown provide a poor match to the distribution of stream members beyond $\gtrsim 70\kpc$ [ref. \cite{belokurov14}]. Crucially, studies using RR Lyrae stars have found stars associated with Sagittarius at distances up to $120$ kpc\cite{hernitschek17}, which indicates that cleaning the sample of Sagittarius stars may be an important step in order to remove possible bias from measurements using the smooth stellar halo. Ideally, one would like to fit the Sagittarius stream {\it and} the smooth halo simultaneously in order to assign membership probabilities that take into account individual measurement errors. Alas, this is a complicated task owing to the poor theoretical understanding of the tidal disruption of the Sagittarius dwarf and its associated stellar stream, one that must be addressed in separate follow-up work\cite{penarrubia11}.

Extended Data Figure~\ref{fig:sgr} shows the angular momentum components of BHBs (blue) and K-Giants (red) located at $r>20\kpc$ within $|B_{\rm sgr}|<20^\circ$ (upper panels) and without $|B_{\rm sgr}|>20^\circ$ (lower panels) from the Sagittarius stream plane\cite{majewski03}. The presence of stream stars in the upper panels are clearly identifiable as an clump of stars around the Sgr dwarf galaxy centroid at $\vec L_{\rm sgr}=(605,-4515,-1267) \kpc\kms$. To remove stream members we define a sphere in angular momentum space with radius $3000\kpc\kms$. Stars that fall in this region are tagged as Sagittarius members and excluded from our sample. This criterion yields 300 BHBs and 562 K~Giants in the smooth halo component, and 20 BHBs and 115 K~Giants in the stream. Using a sphere of 5000 $\kpc\kms$ instead returns 52 BHBs and 217 K~Giants stream members. We have explicitly checked that changing the radius of the angular momentum sphere has a marginal impact on our fits. 

Extended Data Figure \ref{fig:sgr} also serves to mitigate concerns that substructure could affect the results of this work. As the kinematic dipole induced by the disc reflex motion is a leading-order effect in a spherical harmonics decomposition, the fraction of stars in kinematic substructures would need to be comparable to that in the smooth halo in order to alter the mean value as a function of $\ell$ and $b$. After removing Sagittarius stream members, Extended Data Figure~\ref{fig:sgr} shows no evidence for the presence of additional prominent substructures.

\subsection*{The Small Magellanic Cloud}
The Small Magellanic Cloud (SMC) is known to be interacting with the LMC from observations of enhanced star counts in a `bridge' between the two galaxies\cite{besla16}, which suggests that the two galaxies are gravitationally bound to each other despite their high relative speed ($\approx130\kms$, ref. \cite{kallivayalil13}). Observations of the mass enclosed within the luminous radius of the SMC, $M_{\smc}(<3.5\kpc)=2.9\times10^9M_\odot$ (ref. \cite{stanimirovic04}), and the LMC, $M_{\lmc}(<8.7\kpc)=1.7\times10^{10}M_\odot$(ref. \cite{vandermarel14}), 
suggest a mass ratio $M_{\smc}:M_{\lmc}\approx1:10$. Such a low value indicates that, although the SMC may slightly perturb the trajectory plotted in Extended Data Figure~\ref{fig:apex}, its effect on the MW stellar halo can be assumed to subdominant with respect to that of the LMC. 
This work focuses on the LMC, neglecting the presence of its smaller companion. The choice is in part motivated by simplicity, as well as by the results of tests with mock data (see below), which show that the reflex motion of the MW disc is too small to be detected in the kinematics of the outer stellar halo when the infalling satellite models have masses below $M<10^{11}M_\odot$. Given that the total SMC mass may be one order of magnitude below this threshold, $M_{\smc}\sim 10^{10}M_\odot$, we assume that the effect of the SMC on the MW disc can be safely neglected in our analysis.

\subsection*{Milky Way-LMC $n$-body Models} 
We construct live $n$-body models for the MW-LMC pair in order to generate mock-data catalogues, serving as both test our Bayesian inference methods and highlighting the dynamics of the MW-LMC system. 
The MW consists of an exponential disk and a spherical NFW dark matter halo\cite{navarro97} that resemble the measured distributions in the MW\cite{blandhawthorn16}. From these we construct self-consistent distribution functions and generate an equilibrium $n$-body system. The halo and disk mass are set as $M_h=1.6\times10^{12}M_\odot$ and $M_d=4\times10^{10}M_\odot$, respectively. The halo model extends to a virial radius of $r_{\rm virial}=300$ kpc and we quote the halo mass inside this radius.  The halo scale radius is $r_s = 16.6$ kpc. The disk scale length and scale height are $R_d=3$ kpc and $z_0=300$ pc, respectively. We scale the MW so the velocity at the solar circle is $232.8 \kms$ [ref. \cite{mcmillan17}]. The models have a total of 11 million particles: $n_{h}=10^{7}$ and $n_d=10^6$. This numerical resolution was adequate to illustrate the main features of a tidal encounter though future work will increase particle numbers to resolve finer dynamical details of the interaction. The LMC is modeled as a Plummer softened point mass, where the Plummer softening length is fixed at $r_{\rm soft}=12\kpc$. We test three different mass models for the LMC, $M_{\rm LMC}=\{1,2,3\}\times10^{11}M_\odot$. These masses are informed by recent measurements of the LMC mass\cite{penarrubia16,erkal19,fritz19}. One could tune the softening length to match the observed rotation velocity\cite{vandermarel14}; however, for the purposes of our study, the dynamically-relevant quantities are the forces at distances beyond the Galactocentric radius of the LMC ($r>50\kpc$), which dislodge the disc from the outer halo barycentre. For modest changes to the softening length, the forces at $r>50\kpc$ do not change.

We evolved the models using {\sc exp}\cite{weinberg99,petersen20b}, a basis function expansion code that has highly desirable properties for $n$-body simulations, namely well-understood noise properties and high computational efficiency. The basis functions allow us to compute the density and the potential of self-consistent gravitational systems. We track the centre of the MW disc and halo by following an ensemble of particles that have the highest binding energy. We transform all mock data to be in observable heliocentric coordinates. The centre-of-energy of the halo and disc are never displaced from one another by more than 0.5 kpc. The motion of the inner halo is with that of the disc, with the magnitude of the reflex motion smoothly changing with increasing radius\cite{petersen20a}.

From the fiducial spherical MW halo potential, we measure the dynamical times for halo orbits as a function of apocentre, and make predictions for where we expect to see the signal of reflex motion. We theorized in an earlier work\cite{petersen20a} that the magnitude of the reflex motion would linearly grow between 30 and 50 kpc, which is a point in the halo where the dynamical time is approximately equal to the infall time of the LMC. This can be seen in Extended Data Figure~\ref{fig:time}, which shows how the orbital period of halo stars varies as a function of Galactocentric radius. At a given radius, upper and lower limits on the orbital period are given by circular and radial orbits, respectively. We show the estimated timescale for the onset of reflex motion (e.g. when the disc is displaced from the halo barycentre by $|v_{\rm disc}-v_{\rm halo}|>3\kms$) as red lines. While stars near to the LMC may be affected by local perturbations, on average, stellar particles beyond $r\gtrsim 40\kpc$ still bear the signatures of their unperturbed trajectories owing to their long dynamical times.

\subsection*{LMC trajectory}
The relative trajectory of the MW-LMC pair is set by finding an orbit that satisfies the observed constraints. We define the present-day location of the LMC in Cartesian coordinates as $(x,y,z)_{\rm LMC}=(-1.09\pm0.23, -40.61\pm0.45, -27.54\pm0.33)$ kpc and velocities as $(u,v,w)_{\rm LMC}=(-53.5\pm13.1, -216.7\pm7.3,  210.0\pm10.0)\kms$ [refs. \cite{vandermarel02,schonrich10,vandermarel16,mcmillan17,bennett19}]. We call this time $T=0$. We use these coordinates to define a plane for the LMC orbit and compute an analytic trajectory in that plane for the LMC in the spherical MW halo potential. Our fiducial MW model matches the observed enclosed mass profile of the MW\cite{blandhawthorn16}. The LMC orbit is rewound in time to $T=-2.2$ Gyr, at which time the LMC is 450 kpc from the MW centre. We set the LMC to follow a pre-determined analytic first-infall trajectory during the simulation. The pre-determined trajectory does not account for dynamical friction or other distortions in the halo. We list the configurations of the models in Table~\ref{tab:configurations}.

As the Galaxy responds to the LMC infall, the relative phase-space location of the LMC-MW pair at $T=0$ typically differs from the measured values. To alleviate this mismatch we follow an iterative approach. For a given trajectory we first measure the displacement of the disc. We then re-calculate the analytic trajectory using the distance the MW disc travels and repeat the process until a tolerance value is reached. In practice, we end up with a LMC-MW configuration that is slightly different from the measured values, but reproduces the most important facets, namely the pericentre distance and the direction of approach.

We build additional verification mocks by flipping the plane of the LMC trajectory. In these models, the LMC approaches the present-day location with the opposite $v_\ell$. This test allows us to be confident that our Bayesian inference is correctly determining the direction of the apex independently of the actual LMC trajectory. In the flipped models, the LMC is at a smaller radius on average when compared to the standard trajectory, resulting in a slightly higher displacement velocity (Extended Data Figure~\ref{fig:displace}).

As the SMC and LMC are expected to have had multiple close passages\cite{besla12}, the SMC may affect the LMC trajectory. However, given the estimated mass ratio between the two satellites\cite{kallivayalil13}, $M_{\rm SMC}/M_\lmc\sim 1:10$, the influence of the SMC on our constraints is expected to be negligible.

\subsection*{Uniqueness of the model}

\setcounter{table}{0}
\renewcommand{\tablename}{Supplementary Data Table}
\begin{table}
\caption{{\bf Satellite configurations relative to the Galaxy centre.} \label{tab:configurations}}
\begin{center}
\begin{tabular}{|c|c|c|}
\hline
Satellite & $(x,y,z)$ [kpc] & $(v_x,v_y,v_z)~[\kms]$ \\
\hline
LMC &$(-1.09\pm0.23, -40.61\pm0.45, -27.54\pm0.33)$ & $(-53.5\pm13.1, -216.7\pm7.3,  210.0\pm10.0)$\\
\hline
$1\times10^{11}M_\odot$ model &(8.4, -42.6, -18.9)&(-141.4, -215.7, 291.8)\\
$2\times10^{10}M_\odot$ model&(8.6, -42.2, -17.9)&(-152.0, -229.6, 312.5)\\
 $3\times10^{10}M_\odot$ model& (8.3, -41.8, -17.9)&(-163.1, -236.6, 334.3)\\
\hline
\end{tabular}
\end{center}
\end{table}

The observed constraints on the relative position and velocity of the LMC along with the assumption of a first-infall scenario lead to a range of models similar to the one presented in this work depending on assumptions on the initial profile and shape of the MW dark matter halo.

Our mock-production method has three major unknowns: (1) the mass profile of the outer MW dark matter halo (assumed in our model to follow $\rho_{h}\propto r^{-3}$), including the total mass of the MW halo; (2) the mass profile of the LMC; and (3) the effect of dynamical friction on the LMC trajectory (particularly as the LMC trajectory is fixed in our models). The outer mass profile of the MW halo has is still poorly constrained. Worse, they may be biased given that current models do not account for the recent LMC infall\cite{erkal20b}. Our model is also spherical, while the MW halo may not be. The concentration of the MW may affect the magnitude of $v_{\rm travel}$: our models show that the centre of the halo responds with solid body motion in a similar fashion as the disc. At larger Galactocentric radii, the halo gradually transitions to the static approximation that leads to reflex motion. Between these two limits, the halo smoothly transitions from one to the other. Thus, at intermediate radii, the magnitude of the observed reflex motion is a function of the mass distribution in the inner regions of the Galaxy. 
We expect the slope of the mass profile of the LMC to introduce second-order effects on the direction and the magnitude of the MW barycentric motion. 
Finally, it is worth stressing that although the gravitational pull of the LMC affects the motion of halo stars and leads to the formation of a prominent density wake trailing the motion of the satellite\cite{garavitocamargo19}, we ignore the drag force induced by two-body scattering between halo particles and the LMC (a.k.a. dynamical friction) by forcing the latter to move along a pre-determined trajectory. This was done in order to facilitate the search of initial conditions for $n-$body models that match the relative phase-space location of the LMC-MW at present. The effect of dynamical friction depends on the combination of LMC mass and MW potential in a non-trivial way. This is a challenging theoretical problem, which will be studied in follow-up work with the aid of self-consistent $n$-body models of the LMC-MW pair.

Taken together, the simplifications in our $n$-body setup introduce some uncertainty to the trajectory of the LMC. While these models are unlikely to recover the true orbit, our mock catalogues capture the main ingredients of the response of the MW to the LMC infall: the apex direction of the MW disc displacement is set by the LMC trajectory, and the magnitude of the reflex motion is set by the mass of the LMC and the density profile of the MW.

\subsection*{Creating mock catalogues}

In Extended Data Figure~\ref{fig:distances}, we summarize the described data sets (satellites, BHB stars, K Giants, globular clusters, and RR Lyrae stars), which we use to construct mock catalogues with realistic uncertainties. We first select a snapshot of the model at $T=0$, the defined present-day configuration of the MW and LMC. From the dark matter halo particles of the $n$-body snapshot, we define a mock data set that matches the radial distribution of the K Giant sample. We randomly select particles to match the radial distribution requirement by acceptance-rejection and design the mock data set to have the same number of sources as the observed data set.

In the inset of Extended Data Figure~\ref{fig:distances}, we show the measurement of the errors on proper motions and radial velocity as a function of distance for the K Giant, BHB, and RR Lyrae data sets. Using a second-degree polynomial fit, we generate functions for errors on proper motion and radial velocity as a function of distance. We apply these errors to the mock data sets as a function of the dark matter particle Galactocentric distance. Here, we ignore systematic errors in Gaia DR2, which add a mean variation across the sky that is negligible, and only add a small dispersion to the statistical uncertainties of the order of $\epsilon\sim 0.04$ miliarcseconds per year (ref. \cite{lindegren18}). According to Figure~\ref{fig:distances}, the differences introduced by systematic errors in Gaia DR2 can be safely neglected in our analysis when added in quadrature to the proper motion uncertainties of K Giants and BHB stars at $r>40\kpc$. For simplicity, we do not introduce proper motion covariances, as their impact on our analysis is small. We apply a uniform 10\% distance uncertainty to the mock catalog sources.

We model the sky coverage of SEGUE by tracing a region around the high-density regions of K Giants in $(\ell,b)$ space. In practice, this results in near-uniform coverage of the Northern hemisphere ($b>0^\circ$), and sparse coverage of the Southern hemisphere, largely relegated to the quadrant ($\ell<180^\circ,b<0^\circ$). The quadrant hosting the LMC, $(\ell>180^\circ,b<0^\circ)$, is completely unsampled by SEGUE. The discreteness of the SEGUE fields is not modeled, but we do not expect the sampling within the broader SEGUE footprint to bias the measurements.

\subsection*{Tests against mock data catalogues}
We fit mock samples of K Giants drawn from $n$-body models of the LMC-MW pair in order to test the ability of our Bayesian inference analysis to constrain the magnitude and direction of the disc reflex motion. 

Extended Data Figure~\ref{fig:apextraj} shows the location of the apex derived from kinematics of distant halo stars in the SEGUE footprint. Despite the incomplete sky coverage, our fits recover the direction of the relative motion between the disc barycentre and stellar halo particles located at $r>40\kpc$. Because the $n$-body models also provide full information on the position and velocity of dark matter particles, we have checked that the apex direction corresponds to the relative motion between the MW disc and dark matter particles within the radial range $40<r/\kpc<150$ (crossed circles), which may be relevant for directional experiments that aim to detect dark matter on Earth\cite{besla19}. 
As expected, the bounds on the apex direction become more constraining as the LMC mass increases because the reflex motion of the disc becomes stronger. The disc barycentre is currently moving towards a point on its past trajectory across the MW (solid lines) halfway between the current location of the LMC at $(0,0)$ and the point at which it crossed the virial radius of the Galaxy (open circle).
The above results also hold when flipping the infall direction of the LMC, which suggests that the conclusions derived from these mocks are independent of the actual LMC trajectory.

Extended Data Figure~\ref{fig:displace} plots the amplitude of the reflex motion, $v_{\rm travel}$, against the relative velocity between the disc barycentre and dark matter halo particles within $40<r/\kpc<150$ for different LMC masses. This figure reveals a number of interesting points. First, the velocity $v_{\rm travel}$ increases linearly with the LMC mass, as one would expect from simple angular momentum estimates\cite{penarrubia16}. Comparison between solid and dashed lines shows that changing the infall trajectory introduces second-order effects on this relation.  More realistic, non-rigid models of the LMC may create a more complex relation. Second, within statistical uncertainties the amplitude of $v_{\rm travel}$ derived from the kinematics of the stellar halo coincides with the average velocity between the disc barycentre and dark matter halo particles located within the same distance range as the stellar particles. This has interesting implications for follow-up work that studies how $v_{\rm travel}$ varies as a function of Galactocentric distance, which can put constraints on the deformation of the dark matter halo as a response to the gravitational pull of the LMC, and could be potentially used to re-construct the shape of the dark matter halo prior to the LMC infall. Note that the same scale length was assumed in all LMC models, which results in a more concentrated LMC DM profile in the high mass models. This can impact the strength of dynamical friction, which may change the observed velocity of the Galactic disc. Furthermore, the LMC halos in these simulations are static; in a live model the unbound LMC particles become distorted and may impact this linear scaling.

Comparison of Extended Data Figure~\ref{fig:vtravel} with the results plotted in Figure~\ref{fig:displace} puts the total mass of the LMC in the range $M_\lmc/(10^{11} M_\odot)\sim 1$--2 at a 67\% confidence level. This is in good agreement with published estimates\cite{penarrubia16,erkal19,fritz19}. However, the variation of $v_{\rm travel}$ as a function of $M_\lmc$ depends on choices on the MW potential, which makes the mass bounds quoted above model dependent. For example, this calls for follow-up work with live, self-consistent $n$-body models that study the displacement of MW disc in dark matter haloes with different profiles and shapes. Such a study deserves a separate contribution.

In Extended Data Figure~\ref{fig:mockcorner}, we show the results of running mock data sets through our Bayesian inference. When $v_{\rm travel}$ is significantly nonzero (for the $2-3\times10^{11}M_\odot$ models), our inference is well-constrained in all parameters. When $v_{\rm travel}$ is consistent with zero, we are able to constrain the mean velocities and hyperparameters, but are unable to confidently localize the apex of the reflex motion. As the models are built with no rotation, we see that we are able to recover a rotation signal consistent with zero ($\langle v_{\phi}\rangle=0$), even for large values of $v_{\rm travel}$. Interestingly, massive LMC models introduce significant perturbations in the mean motions of halo stars, namely (i) a significant radial compression, $\langle v_r \rangle<0$, and (ii) bulk tangential motion in the polar direction, $\langle v_\theta\rangle >0$. The dynamical response of the stellar halo to both, the LMC gravitational attraction and the sloshing of the MW disc potential, is thus far from trivial and will be explored in separate contributions. Regarding the results presented in this paper, it is worth noting again the modest covariance between the translational motion and the mean bulk velocities $\langle v_r\rangle$,$\langle v_\theta\rangle$ and $\langle v_\phi\rangle$; future all-sky data will help to conclusively break this degeneracy.

\subsection*{Future data sets}

With the dependence of errors on distance and the expected magnitude of the proper motions at $r=40$ kpc in hand, we can forecast what future data sets will require in order to add additional sources. Both BHBs and RR Lyrae are similar absolute magnitudes ($M_{\rm Gaia}\in[0.4,0.7]$, ref. \cite{iorio19}), and would have proper motion errors appropriate for our model if the uncertainties were approximately halved. Future Gaia data releases will likely reach this threshold, greatly expanding the useful data sets. Larger samples of photometrically-determined BHBs are being assembled\cite{starkenburg19} and will benefit from follow-up spectroscopy to develop 6D data sets.
Other bright sources, such as carbon stars\cite{li18,ripoche20} and the more easily-identified subset long-period variables (e.g. Miras\cite{mowlavi18,gaia19}), may also be expanded with future observations.

Finally, a precise identification of halo stars in the solar neighbourhood will enable a dissection of stars with accurate measurements of orbital parameters that may still retain a dynamical memory of earlier points on the barycentre trajectory. Future models at higher resolution will enable studying such stars within the disc volume using mock data sets.

\subsection*{Data Availability}
All data used in this study is publicly available. The data that support the plots within this paper and other findings of this study are available from https://github.com/michael-petersen/ReflexMotion or from the corresponding author upon reasonable request.

\section*{Acknowledgements }

M.S.P. acknowledges funding from UK Science and Technology Facilities Council (STFC) Consolidated Grant and support from Martin Weinberg for usage of the {\sc exp} code.
This work has made use of data from the European Space Agency (ESA) mission
{\it Gaia} (\url{https://www.cosmos.esa.int/gaia}), processed by the {\it Gaia}
Data Processing and Analysis Consortium (DPAC,
\url{https://www.cosmos.esa.int/web/gaia/dpac/consortium}). Funding for the DPAC
has been provided by national institutions, in particular the institutions
participating in the {\it Gaia} Multilateral Agreement.

\end{document}